# Optical activation function using a metamaterial waveguide for an all-optical neural network

YOSHIHIRO HONDA,[1,*] YUYA SHOJI,[2,3] TOMOHIRO AMEMIYA[2,*]

[1] Sony Group Corporation, 4-14-1 Asahi-cho, Atsugi-shi, Kanagawa 243-0014, Japan
[2] Electrical and Electronic Engineering Department, Tokyo Institute of Technology, Meguro, Tokyo 152-8550, Japan
[3] Laboratory for Future Interdisciplinary Research of Science and Technology, Tokyo Institute of Technology, Meguro, Tokyo 152-8550, Japan
* Corresponding authors: Yoshihiro.Honda@sony.com, amemiya.t.ab@m.titech.ac.jp

**In this study, we experimentally demonstrated that the nonlinear optical coefficient of the original Si can be enhanced by incorporating a metamaterial structure into an existing silicon waveguide. The two-photon absorption coefficient enhanced by the metamaterial structure was 424 cm/GW, which is $1.2 \times 10^3$ times higher than that of Si. Using this metamaterial waveguide–based nonlinear optical activation function, we achieved a high inference accuracy of $98.36\%$ in the handwritten character recognition task, comparable to that obtained with the ReLU function as the activation function. Therefore, our approach can contribute to the realization of more power-efficient and compact all-optical neural networks.**

Machine learning, particularly deep neural networks (DNNs), has attracted attention in several fields, such as computer vision, speech recognition, and natural language processing. However, the number of parameters used in DNNs has been increasing in recent years, and latency and power consumption have become significant issues for conventional von Neumann–type electrical processors. Optical neural networks (ONNs), which have the advantages of high speed, low power consumption, and high parallelism, are expected to be the next generation of computing architecture. An ONN is composed of two units: linear matrix-vector multiplication (MVM) and nonlinear activation function (NAF) units (Fig. 1(a)). Several studies have been conducted on the linear MVM unit. This has been realized using Mach–Zehnder interferometer (MZI) meshes [1], micro ring resonator (MRR) weight banks [2], and phase-change material crossbar arrays [3]. Nevertheless, realizing all-optical NAFs is challenging owing to the lack of optical nonlinearity. Therefore, an integration with conventional electronic devices has been proposed for tasks such as activation functions, which are difficult to achieve with optics alone. However, in this approach, the high speed and bandwidth of ONNs are limited by optical–electrical–optical (OEO) conversion.

Optical NAF units have been realized using electro-optical approaches such as electro-absorption modulators [4], semiconductor optical amplifiers [5], semiconductor lasers [6], Si MRRs [7], and MZIs [8], which achieve reconfigurable optical activation functions by voltage control [9]. However, the system becomes more complex owing to the need for external circuits, and the loss due to OEO conversion increases. All-optical NAFs have also been realized using optical approaches such as quantum dots [10] and PhC Fano lasers [11]. These methods can maximize the high-speed capabilities of light, although the issue of complex fabrication process remains.

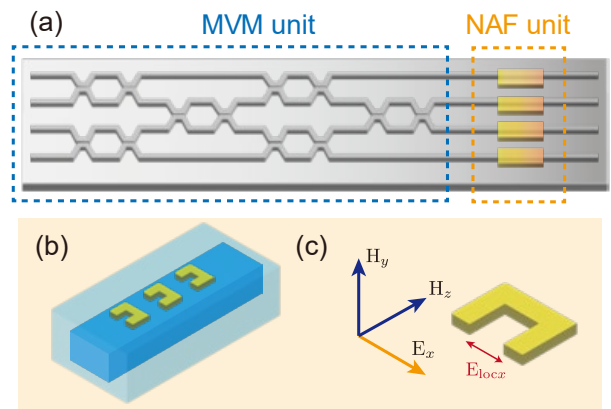

Fig. 1. (a) Single layer of an MZI-ONN with an optical NAF. (b) Schematic of a metamaterial waveguide. (c) An orientation of the SRR with the TE mode.

In this study, we demonstrated that, by incorporating metamaterial structures into Si waveguides (Fig. 1(b)), the nonlinear properties of the material can be enhanced and the ability to function as a NAF for an ONN can be realized. Metamaterials possess multiferroic properties, coupling electric and magnetic fields through their structure, and exhibit significant dispersion near their resonant wavelength, and various applications such as structural color generation have been proposed [12,13]. These properties allow us to achieve slow light effects and nonlinear enhancement that could not be achieved with conventional waveguide structures. To this end, we proposed a new nonlinear optical device. By utilizing the slow light effect of the proposed metamaterial waveguide, the two-photon absorption (TPA) coefficient of Si was increased by $1.2 \times 10^3$ times, resulting in significant nonlinearity. By leveraging the enhanced nonlinearity to implement an NAF, we achieved an inference accuracy comparable with that of conventional digital activation functions in the handwritten digit recognition task (MNIST).

First, we considered the characteristics of light propagation when split-ring resonator (SRR) arrays are placed on an Si waveguide. Fig. 1(c) shows the set orientation of the SRRs to be placed. Assuming the transverse electric (TE) mode as the incident light, the interaction between the magnetic field $H_y$ of the TE mode and the SRR owing to the magnetoelectric (ME) effect of the metamaterial results in the generation of a local electric field $E_x$.

The wave equation at this time can be written as [14]
$$\partial_y \partial_y E_x + (\omega^2(\varepsilon\mu + \zeta^2) - \beta^2)E_x = 0 \quad (1)$$

$$H_z = \frac{-\partial_y}{i\omega\mu} E_x \quad (2)$$

Here, $\varepsilon$, $\mu$, and $\xi$ represent the dielectric permittivity, magnetic permeability, and magnetic-to-electric coupling coefficient, respectively; $\beta$ is the propagation constant along the $z$-direction; and $\omega$ is the angular frequency. The propagation constant $\beta$ derived from Eq. (1) is significantly influenced by the large dispersion characteristics and ME effects near the resonant wavelength of the metamaterial.

Using the nonlinear Schrödinger equation, we considered the time evolution of the slowly varying envelope $A(z,T)$ of the pulse in the waveguide [15].

$$\frac{\partial A}{\partial z} + \frac{\alpha}{2}A + \frac{i\beta_2}{2}\frac{\partial^2 A}{\partial T^2} - \frac{\beta_3}{6}\frac{\partial^3 A}{\partial T^3}$$
$$= i\gamma(|A|^2 A) - \frac{\beta_{TPA}}{2A_{eff}}|A|^2 A$$
$$- N_c\left(\frac{\sigma}{2} + \frac{2k_c i\pi}{\lambda_0}\right)A \quad (3)$$

Near the resonant wavelength of the metamaterial waveguide, the propagation constant $\beta$ and the modal group velocity $S$ changes accordingly as shown in Eq. (1). Here, $S = n_g/n_{Si}$, where $n_g$ is the group refractive index of the metamaterial waveguide and $n_{Si}$ is the refractive index of the Si waveguide. In this case, $\alpha, \gamma, \beta_{TPA}, \sigma$, and $k_c$ can be replaced with $\alpha \times S$, $\gamma \times S^2$, $\beta_{TPA} \times S^2$, $\sigma \times S$, and $k_c \times S$, respectively, when the group velocity $S$ is introduced [16]. We conducted an experiment to verify the enhancement effect of the nonlinear absorption coefficient owing to TPA by placing metamaterials on an optical waveguide.

The fabrication process of the metamaterial waveguide is illustrated in Fig. 2. An Si-on-insulator substrate, with a 220 nm (thickness) × 500 nm (width) Si waveguide, was used as the initial wafer. The SiO₂ overcladding layer thickness was 1.5 µm; in the area where the metamaterial structure was installed, the overcladding thickness was reduced through etching. The overcladding layer thickness after thinning had a significant effect on the strength of the interference between the optical modes propagating in the waveguide and metamaterial, and consequently on the nonlinearity strength and propagation loss. In this study, the overcladding thickness was set to 85 nm, considering the tradeoff between the nonlinearity strength and propagation loss.

Next, a Ti–Au SRR array was fabricated using polymethyl methacrylate (PMMA) coating, electron beam lithography, electron beam deposition, and a lift-off process. This fabrication method made it possible to integrate metamaterials in the back-end process without disrupting the CMOS-compatible silicon photonics process. In other words, it becomes easier to integrate with other optical elements fabricated such as MZIs using silicon photonics processes [17].

Figure 3 shows the metamaterial waveguide fabricated using this method. The fabricated SRRs were not arranged in a single array at the center of the Si waveguide but in multiple arrays. This arrangement allowed the adjacent SRRs to interact with the Si waveguide core, which is expected to increase the nonlinearity. Various orientations are available for arranging the SRRs on the waveguide. We considered an arrangement pattern (Fig. 1(c)) in which only the magnetic field $H$ resonates with the SRR when the TE mode is incident.

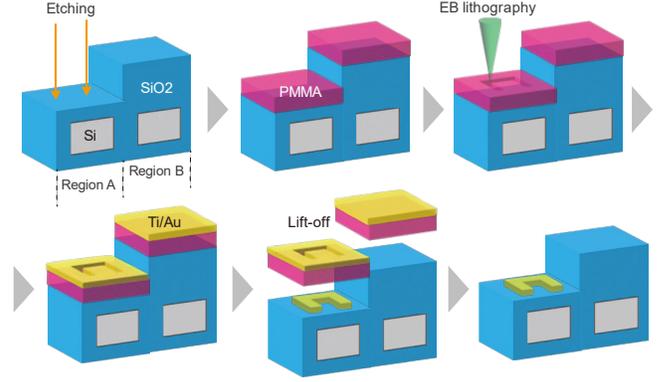

Fig. 2. Fabrication process of the metamaterial waveguide.

First, we evaluated the characteristics of SRRs fabricated on the waveguide and measured the resonant wavelength using a Fourier transform infrared spectrometer (FTIR) when the structural parameters of SRR were varied as $w = 130, 150, 170$ nm (other structural parameters were maintained constant). In FTIR measurements, a plane wave was incident such that SRRs resonated with the electric field $E$. The spectral results for each parameter are shown in Fig. 3, which showed that the SRR with $w = 150$ nm had a resonant wavelength of approximately 1.59 µm. The reason for the low transmittance in FTIR measurements is attributed to the filling factor of SRRs. The measurement area was 100 µm × 100 µm; therefore, the transmission rate was affected by the area occupied by SRRs.

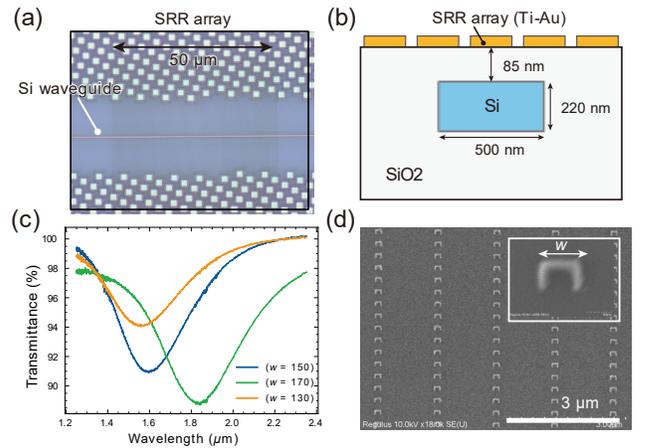

Fig. 3. (a) Optical micrograph of SRR arrays on the Si waveguide. (b) Cross section of the waveguide and the metamaterial array. (c) FTIR transmission spectra of SRRs with different sizes. (d) Scanning electron microscope image of the SRR.

Figure 4 (a) shows the experimental setup. The optical nonlinear characteristics of the metamaterial optical waveguide were determined by injecting ultrashort pulses into the submicron-sized waveguide and measuring the output as a function of the input power. The light source produced a 4-ps pulse with a wavelength of 1.59 μm and a repetition rate of 25 MHz. The input power, controlled by a variable attenuator, passed through a polarizer and a beam splitter and was coupled with the waveguide in free space using an objective lens with a numerical aperture (NA) of 0.25 and a focal length of 12 mm. The polarizer was adjusted to excite the TE mode in the waveguide. Light was introduced into the waveguide via free-space coupling to prevent the adverse effect of the nonlinearity induced in the fiber on the transmission spectrum of the waveguide. The output from the chip was collected using a tapered lensed fiber and measured using an optical spectrum analyzer (OSA) and a power monitor. We used the broadband amplified spontaneous emission (ASE) source to measure the spectrum of the metamaterial waveguide. To achieve low coupling loss with the optical fiber, a beam spot size converter comprising a 100-mm-long tapered core was used at both ends. The total length of the waveguide, including the spot size converter, was 5 mm.

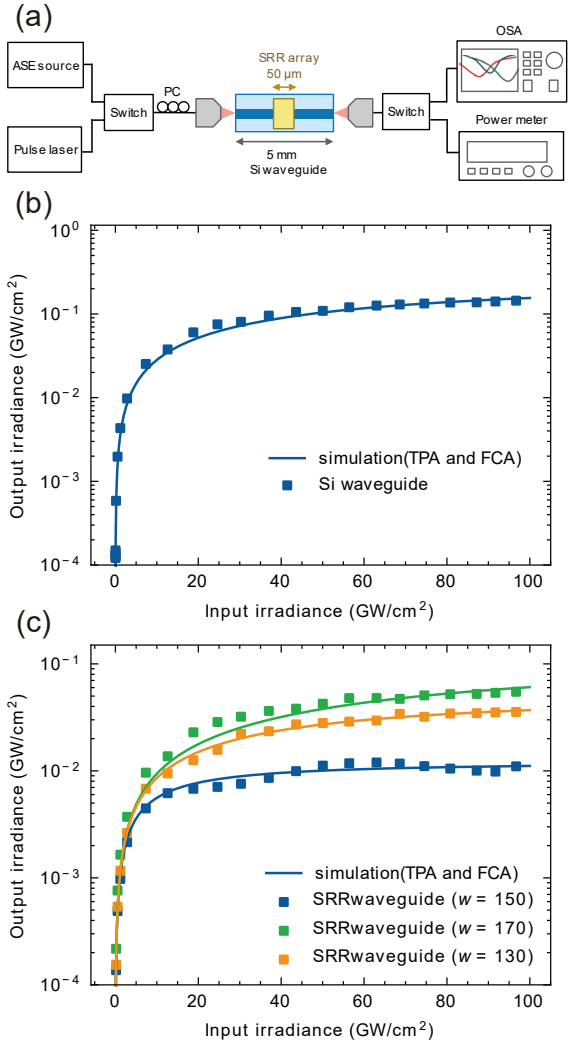

Fig. 4. (a) Experimental setup. Experimental results for the (b) silicon and (c) metamaterial waveguides with different SRR sizes.

Figure 4(b) shows the input–output characteristics of the Si waveguide measured in the experiment, and Fig. 4(c) shows those of the metamaterial waveguide. The horizontal and vertical axes represent the input and output powers, respectively. We compared the original Si waveguide shown in Fig. 4(b) with the metamaterial waveguides having different resonant frequencies, as shown in Fig. 4(c). The results showed that the metamaterial waveguide with SRRs ($w = 150$), where the center wavelength of the incident light matched the resonant wavelength, exhibited output saturation at the lowest power of approximately 5 mW. In other words, the nonlinearity of the Si waveguide can be enhanced by matching the resonant wavelength of the SRR with that of the incident light.

Next, we quantitatively compared the rate of enhancement of the nonlinearity of the fabricated metamaterial waveguide with that of the original Si waveguide. The saturation phenomenon in the waveguide is dominated by the TPA and free-carrier absorption (FCA) [18]. Here, we used the equation for nonlinear absorption (Eq. 4), which includes parameters such as the propagation loss $\alpha_l$, TPA coefficient $\beta_{\text{TPA}}$, and FCA coefficient $\sigma N_c$. By fitting the experimental results (Fig. 4) using parameters obtained from the experimental conditions, we estimated the TPA coefficient as follows:

$$\frac{dI}{dz} = -\alpha_l I - \beta_{\text{TPA}} I^2 - \sigma N_c I \quad (4)$$

The propagation loss of the Si waveguide, FCA cross section $\sigma$, and free carrier density were set to 2.5 dB/cm, $1.45 \times 10^{-17}$ cm$^2$, and $N_c = \frac{\beta_{\text{TPA}} \sqrt{\pi} T I^2}{4\hbar\omega}$, respectively. Using these parameters, the TPA coefficient of the Si waveguide was estimated to be $\beta_{\text{TPA}} = 0.34$ cm/GW as a result of fitting; this value is consistent with the absorption coefficient of Si reported in the literature [19]. Similarly, the TPA coefficients for each metamaterial waveguide (50 μm) were as follows: 424 cm/GW for $w = 150$, 62.7 cm/GW for $w = 130$, and 3.03 cm/GW for $w = 170$. These results indicate that, by arranging the metamaterial structure appropriately, the TPA coefficient of Si can be enhanced by up to $1.2 \times 10^3$ times. As mentioned above, this enhancement of nonlinearity was achieved by effectively reducing the propagation constant $\beta$. The measurement results confirmed that the metamaterial waveguide exhibited a very strong slow light effect with a group refractive index of $n_g \approx 140$ ($n_g \approx 4$ for an Si waveguide).

Next, we used the nonlinearity enhanced by metamaterials to confirm whether the proposed metamaterial waveguide can function as an optical NAF for ONNs. We replaced the experimentally obtained nonlinear functions of each waveguide with NAFs, such as ReLU, and conducted training and inference evaluations for the MNIST task. The results of the recognition accuracy when using the convolution neural network (CNN) model and the nonlinearity of each waveguide as NAFs are shown in Fig. 5. The results indicated that using a metamaterial waveguide, whose resonant wavelength matches the wavelength of the incident light, as an NAF can improve the recognition accuracy to 98.36%. The comparison results of the fabricated waveguide with those of other metamaterial waveguides, Si waveguides, and the ReLU function are presented in Table 1, which show that using a metamaterial waveguide with $w = 150$ achieves the same accuracy as the commonly used ReLU function. Furthermore, the recognition accuracy of other metamaterial waveguides ($w = 130, 170$) is similar to that of Si waveguides. This is because, both the nonlinearity and the propagation loss of the waveguide increased; consequently, the difference between the other metamaterial and Si waveguides was not significant when considered as an NAF. Therefore, the optical activation function of the metamaterial waveguide proposed in this study can operate at a lower power (∼5 mW) and with a smaller interaction length of 50 μm, as shown in Fig. 4. Furthermore, it is possible to fabricate an NAF combined with an MVM unit comprising an MZI in the ONN system.

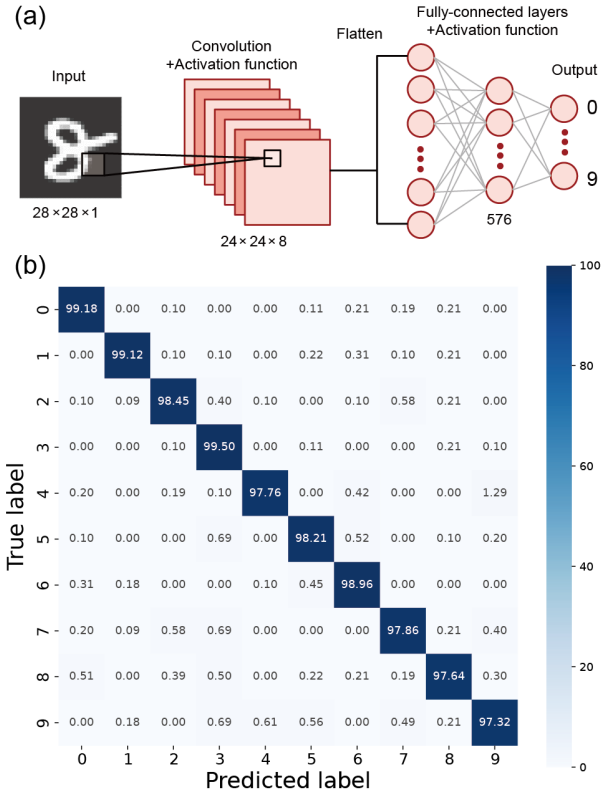

Fig. 5. (a) Architecture of the convolution neural network. (b) Inference results using the optical activation function.

Hence, for the first time, we experimentally demonstrated that, by arranging metamaterials on Si waveguides and utilizing both ME effect and the resonant characteristics of the metamaterials, a significant slow light effect ($n_g \approx 140$) can be achieved. We also showed a substantial enhancement in nonlinearity, with a TPA coefficient $1.2 \times 10^3$ times that of the Si waveguide. Moreover, by using the nonlinearity of this metamaterial waveguide as the optical NAF, we achieve a recognition accuracy of 98.36% in MNIST. The proposed optical NAF can be fabricated in the backend process while maintaining CMOS-compatible processes, contributing to the integration with the MZI–MVM unit and the realization of more power-efficient and compact all-optical neural networks.

Table 1. Test Accuracy of the CNN with Variable NAFs

| Activation function | Test accuracy |
|---|---|
| SRR waveguide (w=150) | 98.36% |
| SRR waveguide (w=170) | 96.28% |
| SRR waveguide (w=130) | 96.77% |
| Si waveguide | 96.48% |
| ReLU | 98.38% |

**Funding.** Japan Society for the Promotion of Science (#22H01520); Adaptable and Seamless Technology Transfer Program through Target-Driven R&D (JPMJTR22RG); Adopting Sustainable Partnerships for Innovative Research Ecosystem.

**Disclosures.** The authors declare that there are no conflicts of interest related to this article.

**Data availability.** Data underlying the results presented in this paper are not publicly available at this time but may be obtained from the authors upon reasonable request.

**References**

1. Y. Shen, N. C. Harris, S. Skirlo, *et al.*, Nat. Photonics **11**, 441–446 (2017).
2. A. N. Tait, T. F. de Lima, M. A. Nahmias, *et al.*, IEEE Photonics Technol. Lett. **28**, 887–890 (2016).
3. J. Feldmann, N. Youngblood, M. Karpov, *et al.*, Nature **589**, 52–58 (2021).
4. R. Amin, J. K. George, S. Sun, *et al.*, APL Mater. **7**, 081112 (2019).
5. G. Mourgias-Alexandris, A. Tsakyridis, N. Passalis, *et al.*, Opt. Express **27**, 9620–9630 (2019).
6. J. Crnjanski, M. Krstić, A. Totović, *et al.*, Opt. Lett. **46**, 2003–2006 (2021).
7. A. N. Tait, T. Ferreira de Lima, M. A. Nahmias, *et al.*, Phys. Rev. Appl. **11**, 064043 (2019).
8. M. M. Pour Fard, I. A. D. Williamson, M. Edwards, *et al.*, Opt. Express **28**, 12138–12148 (2020).
9. A. Jha, C. Huang, and P. R. Prucnal, Opt. Lett. **45**, 4819–4822 (2020).
10. M. Miscuglio, A. Mehrabian, Z. Hu, *et al.*, Opt. Mater. Express **8**, 3851 (2018).
11. T. S. Rasmussen, Y. Yu, and J. Mork, Opt. Lett. **45**, 3844–3847 (2020).
12. M. Miyata, H. Hatada, and J. Takahara, Nano Lett. 16, 3166–3172 (2016).
13. A. M. Urbas, Z. Jacob, L. D. Negro, N. Engheta, *et al.*, J. Opt. **18**, 093005 (2016).
14. Y. Honda, E. Igarashi, Y. Shoji, *et al.*, Opt. Express **31**, 32017–32043 (2023).
15. L. Yin and G. P. Agrawal, Opt. Lett. **32**, 2031–2033 (2007).
16. C. Monat, B. Corcoran, M. Ebnali-Heidari, *et al.*, Opt. Express **17**, 2944–2953 (2009).
17. Y. Zhang, Q. Du, C. Wang, *et al.*, Optica 6, 473 (2019).
18. H. Yamada, Ã, M. Shirane, T. Chu, *et al.*, Jpn. J. Appl. Phys. **44**, 6541–6545 (2005).
19. A. D. Bristow, N. Rotenberg, and H. M. van Driel, Appl. Phys. Lett. **90**, 191104 (2007).